\documentclass{bmcart}

\usepackage[utf8]{inputenc} %unicode support

\usepackage{mathptmx}       % selects Times Roman as basic font
\usepackage{helvet}         % selects Helvetica as sans-serif font
\usepackage{courier}        % selects Courier as typewriter font
\usepackage{type1cm}        % activate if the above 3 fonts are not available on your system
\usepackage{makeidx}         % allows index generation
\usepackage{graphicx}        % standard LaTeX graphics tool when including figure files
\usepackage{subcaption}
\usepackage{multicol}        % used for the two-column index
\usepackage[bottom]{footmisc}% places footnotes at page bottom
\usepackage{amsfonts}
\usepackage[cmex10]{amsmath}

\startlocaldefs
\endlocaldefs

\begin{document}

%%% Start of article front matter
\begin{frontmatter}

\begin{fmbox}
\dochead{Research}


\title{Optimal shattering  of complex networks}

%%%%%%%%%%%%%%%%%%%%%%%%%%%%%%%%%%%%%%%%%%%%%%
%%                                          %%
%% Enter the authors here                   %%
%%                                          %%
%% Specify information, if available,       %%
%% in the form:                             %%
%%   <key>={<id1>,<id2>}                    %%
%%   <key>=                                 %%
%% Comment or delete the keys which are     %%
%% not used. Repeat \author command as much %%
%% as required.                             %%
%%                                          %%
%%%%%%%%%%%%%%%%%%%%%%%%%%%%%%%%%%%%%%%%%%%%%%

\author[
   addressref={aff1},                   % id's of addresses, e.g. {aff1,aff2}
   email={nicolebalashov@gmail.com}   % email address
]{\inits{N}\fnm{Nicole} \snm{Balashov}}
\author[
   addressref={aff1},
   email={reuven@math.biu.ac.il}
]{\inits{R}\fnm{Reuven} \snm{Cohen}}
\author[
   addressref={aff1},                   % id's of addresses, e.g. {aff1,aff2}
   email={avieli@gmail.com}   % email address
]{\inits{A}\fnm{Avieli} \snm{Haber}}
\author[
   addressref={aff2},                   % id's of addresses, e.g. {aff1,aff2}
   email={krivelev@tau.ac.il}   % email address
]{\inits{M}\fnm{Michael} \snm{Krivelevich}}
\author[
   addressref={aff1},                   % id's of addresses, e.g. {aff1,aff2}
   corref={aff1},                       % id of corresponding address, if any
   email={simi.haber@biu.ac.il}   % email address
]{\inits{S}\fnm{Simi} \snm{Haber}}

%%%%%%%%%%%%%%%%%%%%%%%%%%%%%%%%%%%%%%%%%%%%%%
%%                                          %%
%% Enter the authors' addresses here        %%
%%                                          %%
%% Repeat \address commands as much as      %%
%% required.                                %%
%%                                          %%
%%%%%%%%%%%%%%%%%%%%%%%%%%%%%%%%%%%%%%%%%%%%%%

\address[id=aff1]{%                           % unique id
  \orgname{Department of Mathematics, Bar-Ilan University}, % university, etc
  %\street{Waterloo Road},                     %
  \postcode{5290002}                                % post or zip code
  \city{Ramat-Gan},                              % city
  \cny{Israel}                                    % country
}
\address[id=aff2]{%
  \orgname{Department of Mathematics, Tel-Aviv University},
  %\street{D\"{u}sternbrooker Weg 20},
  \postcode{6997801}
  \city{Tel-Aviv},
  \cny{Israel}
}

%%%%%%%%%%%%%%%%%%%%%%%%%%%%%%%%%%%%%%%%%%%%%%
%%                                          %%
%% Enter short notes here                   %%
%%                                          %%
%% Short notes will be after addresses      %%
%% on first page.                           %%
%%                                          %%
%%%%%%%%%%%%%%%%%%%%%%%%%%%%%%%%%%%%%%%%%%%%%%

%\note{Sample of title note}     % note to the article
%\note[id=n1]{Equal contributor} % note, connected to author

\end{fmbox}% comment this for two column layout

%%%%%%%%%%%%%%%%%%%%%%%%%%%%%%%%%%%%%%%%%%%%%%
%%                                          %%
%% The Abstract begins here                 %%
%%                                          %%
%% Please refer to the Instructions for     %%
%% authors on http://www.biomedcentral.com  %%
%% and include the section headings         %%
%% accordingly for your article type.       %%
%%                                          %%
%%%%%%%%%%%%%%%%%%%%%%%%%%%%%%%%%%%%%%%%%%%%%%

\begin{abstractbox}

\begin{abstract} % abstract
We consider optimal attacks or immunization schemes on different models of random graphs. We derive bounds for the minimum number of nodes needed to be removed from a network such that all remaining components are fragments of negligible size.

We obtain bounds for different regimes of random regular graphs, Erd\H{o}s-R\'enyi random graphs, and scale free networks, some of which are tight. We show that the performance of attacks by degree is bounded away from optimality.

Finally we present a polynomial time attack algorithm and prove its optimal performance in certain cases.

\end{abstract}

%%%%%%%%%%%%%%%%%%%%%%%%%%%%%%%%%%%%%%%%%%%%%%
%%                                          %%
%% The keywords begin here                  %%
%%                                          %%
%% Put each keyword in separate \kwd{}.     %%
%%                                          %%
%%%%%%%%%%%%%%%%%%%%%%%%%%%%%%%%%%%%%%%%%%%%%%


% MSC classifications codes, if any
%\begin{keyword}[class=AMS]
%\kwd[Primary ]{}
%\kwd{}
%\kwd[; secondary ]{}
%\end{keyword}

\end{abstractbox}
%
%\end{fmbox}% uncomment this for twcolumn layout

\end{frontmatter}

%
%\titlerunning{Optimal attacks}  % abbreviated title (for running head)

%
\section{Introduction}
One of the most
studied questions in complex networks is the resilience of networks under different
failure models and attack strategies \cite{A:albert_&_jeong_&_barabasi2000,A:cohen_&_erez_&_ben-avraham_&_havlin2000,A:callaway_&_newman_&_strogatz_&_watts2000}. In particular, one
wishes to know the optimal attack strategy that will lead to
fragmentation by removal of a minimal fraction of the nodes.
This information is important for estimating the vulnerability
of network infrastructures, and also for devising optimal immunization
strategies for populations and computer networks.

The main methods that have been proposed for targeted attacks
on networks via node removal have been based on attack by
highest degree \cite{A:albert_&_jeong_&_barabasi2000,A:callaway_&_newman_&_strogatz_&_watts2000,A:cohen_&_erez_&_ben-avraham_&_havlin2001},
and attack by highest betweenness
centrality \cite{A:magoni2003}. Some methods based on more advanced
algorithms for graph partitioning have also been proposed
\cite{A:paul_&_cohen_&_sreenivasan_&_havlin_&_stanley2007}, and led to improved upper bounds
on the minimal fraction of nodes that should be removed to
shatter a networks.

In recent years, several works studied optimal attacks on networks, presenting sophisticated, highly efficient algorithms for choosing minimal sets of nodes whose removal leads to complete fragmentation of the network. In \cite{Braunstein12368} an efficient dismantling method is presented, using the replica method on the generating function. In \cite{Ren6554} efficient dismantling is considered when costs are attributed to the different nodes. In \cite{A:morone_&_makse2015} finding sets of influential nodes is considered using the cavity and Extremal Optimization (EO) methods.
In \cite{PhysRevE.94.012305} belief propagation is used  to find efficient shattering sets.
In \cite{A:osat_&_faqeeh_&_radicchi2017} optimal attacks on multiplex networks are studied using simulated annealing methods.

As for measuring the resilience of a network to shocks, there are again quite a few papers. The following is far from being a comprehensive survey. In \cite{CERQUETI2019320} a shock is injected on one node and the resilience of the (weighted) network is measured by its ability to absorb this shock. In \cite{PhysRevE.91.022805} network robustness
is evaluated in terms of the ability to identify the attack prior to network disruption. The affect of nodes removal on the diameter of the network, a natural parameter, was studied in \cite{doi:10.1504/IJCEE.2018.088314}. Both of these studies regard \emph{attack by degree} as representative for an intentional attack. An important breakthrough was made in \cite{Gao2016UniversalRP}, where the \emph{dynamics} of the system was accounted for, in addition to network topology. One may ask about the resilience of the community structure in a network, and indeed this approach was taken in \cite{RAMIREZMARQUEZ2018466}, and tested against link removal. For a review of definitions and measures of system resilience we refer the reader to \cite{HOSSEINI201647}.

In this work, we present results for several network classes, giving upper and lower bounds on the size of the minimal set of nodes to be removed in order to shatter (or dismantle) a network. We first survey the method of generating functions and results on random attacks and attack by degree, in order to show that random attack and attack by degree are not asymptotically optimal for any class of generalized (configuration model) random graphs. We then present exact results and bounds on the size of the shattering set for various random graph classes. Eventually, we present a polynomial time algorithm for efficient shattering of random graphs. We show, using exact methods, that for 3-regular graphs our algorithm obtains asymptotically optimal results. The performance of this algorithm for other classes of random graphs remains an open question.

\section{Random Attacks}
The problem of removing nodes in order to shatter the network into small pieces is of importance both in order to determine the resilience of a network to various attacks and in order to immunize a network (for example, a social network) against the spreading of an epidemic disease. By ``shattering'' a network, we mean breaking the network into small components, each of which having size $o(n)$, where $n$ is the number of nodes. That is, breaking the network into pieces whose sizes are negligible compared to the original network, by means of nodes removal.

The simplest attack mechanism on a network is removing nodes uniformly at random. This is the standard percolation model. In order to study the effect of this attack on the network, one can employ the generating function method (See, e.g., \cite{A:callaway_&_newman_&_strogatz_&_watts2000}). We consider the configuration model, where no correlations exist between neighbouring nodes. Given a degree distribution $P(k)$, and probability $p(k)$ for a degree $k$ node to exist (i.e., not to be deleted), one can write the generating function for this distribution as
\begin{equation}
F_0(x)=\sum_{k=0}^\infty P(k)p(k)x^k\;.
\end{equation}
Similarly, one can write the generating function for the reciprocal degree distribution of a node reached by following an edge (i.e., the degree of the node disregarding the edge through which it was reached):
\begin{equation}
F_1(x)=\frac{F_0'(x)}{\langle k\rangle}=\frac{\sum_{k=0}^\infty kP(k)p(k)x^{k-1}}{\langle k\rangle}\;,
\end{equation}
where ${\langle k\rangle}$ is the average degree.
The generating function for branch sizes reached by following a random edge is given by the recursive equation.
\begin{equation}
H_1(x)=1-F_1(1)+xF_1(H_1(x))\;,
\end{equation}
where $F_1(1)$ is the probability of a node reached by following a random edge to exist.
and the generating function for the probability of a node to belong to a component of some finite size  is given by
\begin{equation}
H_0(x)=1-F_0(1)+xF_0(H_1(x))\;.
\end{equation}
$H_0(1)$ is the normalization of $H_0(x)$. It may happen that $H_0(1)=1$, in which case all components are finite and no giant component exists, or that $H_0(x)<1$ in which case, a giant component exists and contains a fraction $1-H_0(1)=1-F_0(1)+xF_0(u)$ of the nodes, where $u$ is the solution of the self consistent equation
\begin{equation}
\label{eq:u}
u=1-F_1(1)+F_1(u)\;.
\end{equation}
One can observe that such a solution exists only if $F_1'(1)>1$.

For random attacks $p(k)=p:=1-q$ is independent of $k$ and using
Eq.\ (\ref{eq:u}) one can deduce that the criterion for the existence of a giant component is that there exist a solution $u<1$, which exists only if
\begin{equation}
p>p_c:=\frac {1}{\kappa-1}\;,
\end{equation}
where $\kappa:=\frac{\langle k^2\rangle}{\langle k\rangle}$ is the ratio of the first two moments of the distribution \cite{A:cohen_&_erez_&_ben-avraham_&_havlin2000}.
Specifically, using the moments of the constant and Poisson distributions, respectively, one can deduce that the percolation thresholds for the random $d$-regular and Erd\H{o}s-R\'enyi networks are $\frac{1}{d-1}$ and $\frac{1}{\langle k\rangle}$.

Another result that can be deduced using this formalism is that the probability of a (non-deleted) node of degree $k$ to belong to a finite component is the probability that all of the branches emanating from it are finite, i.e., the probability is $u^k$.

\section{Targeted Attack Strategies}
Naturally, random attacks are not expected to give optimal results. Indeed, a simple and more effective attack strategy is starting the removal with the high degree nodes, as they play a more substantial role in the network connectivity. Using a function
\begin{equation}
q(k)=
\begin{cases}
1 & k<k_0\\
0 & k>k_0\\
\alpha & k=k_0
\end{cases}\;,
\end{equation}
for some $k_0$ and $\alpha$. Solving for the values of  $k_0$ and $\alpha$ that give criticality, one can find the critical fraction for removal.

The attack by degree strategy is, however, suboptimal. This can be seen from the fact that for any finite $k$, the probability that a removed node of degree $k$ does not even belong to the giant component, and therefore its removal is unbeneficial, is $u^k$. Therefore, a finite fraction of the removed nodes are not included in the giant component to begin with, and thus the method is not even asymptotically optimal. Furthermore, for random regular networks, targeted attack by degree is completely equivalent to random removal.

In order to develop a better attack strategy one may consider methods such as adaptive attack by degree or attack by betweenness centrality. However, these strategies are very hard to analyse.

\section{Bounds on optimal attacks}
We define $c_f$ as the minimal fraction of nodes that
are to be removed before the network is shattered (i.e., becomes fragmented into sublinear components).
It is clear that for any network $c_f\leq q_c=1-p_c$,
where $p_c$ is the percolation threshold for random removal
of nodes. For a $d$-regular graph (where each node has degree
$d$) this yields the upper bound
\[c_f \leq 1-\frac{1}{d-1}=\frac{d-2}{d-1} . \]

However, one can ask the question: Given complete knowledge of the network, and unlimited computational power, what is the smallest fraction of the nodes that can be removed in order to break the network into small pieces, each of size $o(n)$? One might be tempted to consider the possibility of shattering the network by removing only a zero fraction, $o(1)$ of the nodes. Indeed, in the case of a square grid, for example, which is a regular network with fixed degree 4, one can remove $n^{1/4}=o(n)$ equally spaced rows and columns and shatter the network into pieces of size $O(\sqrt{n})$.

For some cases, such as random regular networks, this can be shown to be impossible. Indeed, improved bounds were obtained in \cite{A:edwards_&_farr2001,A:edwards_&_farr2008}:
\[ \frac{d-2}{2d-2} \leq c_f \leq \frac{d-2}{d+1} .\]
In this paper we establish better lower and upper bounds
on $c_f$, focusing on sparse random graphs. In particular for random regular graphs we show
\begin{equation} \label{Eq:bounding_by_ind_number}
1-2\frac{\alpha(G)}{n} \leq c_f \leq 1-\frac{\alpha(G)}{n}\;,
\end{equation}
where $\alpha(G)$ is the independence number of $G$. When $d$ is large enough the independence number is known \cite{A:frieze_&_luczak1992} to satisfy $\alpha(G) \approx 2n\ln d/d$. In this case we obtain
\[ c_f \approx 1-\frac{\alpha(G)}{n} .\]
We provide matching results for Erd\H{o}s-R\'enyi random graphs.

\section{Results}
For the analytical results, we mainly exploit structural graph properties such as expansion, domination number and independence and obtain deterministic connections to the shattering number. We then apply known estimations of these parameters for random graphs either directly or via contiguity arguments.

As an example we establish a lower bound on the minimal fraction of nodes needed to be removed in order to shatter a regular network into small
components of size  $o(n)$.
Consider shattering a graph to $m$ disjoint clusters  $C_i$,
$i=1,\ldots,m$ by
deleting a set of $|S|=c_f n$ nodes. $c_f$ is therefore the
fraction of removed nodes. Notice that
$\sum_{i=1}^m |C_i|+ |S|=n$. Thus,
$\frac{1}{n}\sum_i |C_i|+c_f=1$. Denote by $B_i$ the nodes
on the boundary of $C_i$,
i.e., the neighbors of nodes of cluster $i$ outside of the cluster. Since
the clusters are disconnected for all $i,j$, all boundary nodes
for any cluster $i$ must be removed.
Therefore, for every $i$, all nodes in $B_i$ are deleted.
Now, in a random regular graph with high probability each cluster is locally almost tree like. Therefore, up to an additive constant,
$|B_i|=(d-1)|C_i|-|C_i|=(d-2)|C_i|$.
Since every node has exactly $d$ neighbors, a node can not participate in
more than $d$ of the $B_i$s. Thus $\frac{1}{d}\sum_i |B_i|\leq |S|$.
Therefore,
\begin{equation}
n\geq \sum_i C_i +\frac{d-2}{d}\sum_i C_i
=\frac{2d-2}{d}\sum_i C_i.
\end{equation}
Summarizing the above we get:
\begin{equation}
\label{eq:expansion_bound}
c_f=1-\frac{1}{n}\sum_i C_i\geq \frac{d-2}{2d-2}.
\end{equation}

This approach can be shown to give asymptotically
tight solution for $d=3$ (i.e., it matches the upper bound shown in
\cite{A:edwards_&_farr2001}). However, for large values of $d$ it deviates
considerably from the exact value. Indeed, for $d\to\infty$,
Eq.\ \eqref{eq:expansion_bound} leads to $c_f\geq \frac{1}{2}$, where
as will be shown below $c_f\to 1$.

In order to give a better lower bound on $c_f$ for random regular graphs with large constant degree we observe the following: Random regular graphs are locally tree like, having a bounded number of short cycles. Therefore after the network is shattered we expect the remaining components to be trees. A tree can be shattered into isolated vertices by removing at most half of its nodes.
Therefore, it can be deduced that the number of nodes remaining after the attack
is at most twice the size of the largest independent set. Since removing all but an independent set clearly shatters the graph, we obtain Equation \eqref{Eq:bounding_by_ind_number}. With some further effort we can improve the lower bound to close the gap and match the upper bound. The same approach may be applied to the Erd\H{o}s-R\'enyi model $G(N,p=c/N)$ where we get
\begin{equation}
c_f\approx 1-\frac{\log c}{c}.
\end{equation}

We discuss theoretical and practical implications of our results in Section \ref{Sec: Summary}.

\section{Shattering Scale Free Networks}
Targeted attack strategies usually begin by attacking the hubs of the scale free networks. This is based on the idea that the hubs are the glue holding the network together, due to their high degree and large number of neighbours. Indeed, one may use this intuition to obtains bounds on the hardness of shattering a scale free networks.

Consider a scale free network with degree distribution
\begin{equation}
P(k)=\frac {k^{-\gamma}}{\zeta(\gamma)}\;,
\end{equation}
where $\zeta(\gamma)=\sum_{k=1}^\infty k^{-\gamma}$. This network can be shattered using the following two stage process:
\begin{enumerate}
\item Remove all nodes above degree $d$. This contains a fraction of $\sum_{k=d+1}^\infty k^{-\gamma}/\zeta(\gamma)$ of the nodes.
\item \label{Irem:upper bound}Shatter the remaining network, requiring at most $\frac{d-2}{d+1}$ of the network.
\end{enumerate}
One can choose $d$ as to minimize the sum of these two terms in order to obtain an upper bound on the size of the required shattering set. In fact, a better bound can be obtained by noticing that after the removal of the hubs, the remaining nodes form a random network with degree distribution
\begin{equation}
\tilde P(k)=\sum_{c=k}^\infty \frac {k^{-\gamma}}{\zeta(\gamma)}\binom{c}{k}p^k (1-p)^{c-k}\;,
\end{equation}
with
\begin{equation}
p=\sum_1^{d} \frac{kP(k)}{\langle k\rangle}
\end{equation}
denoting the probability of an edge to lead to an undeleted node (i.e. the fraction of edges leading to undeleted nodes). If one can bound the size of the shattering set for a random graph with this degree distribution, a better bound on the overall shattering set size can be obtained. In particular, by showing that the remaining graph is close enough to random graph in the sense of having all local neighborhoods as almost trees with high probability, one may replace the upper bound $\frac{d-2}{d+1}$ in Point \ref{Irem:upper bound} above by $1-\alpha(G)/n$.

\section{Algorithmic aspects}
Finding an optimal shattering set is NP-hard, but when the input is a random graph the problem becomes tractable. We propose the following algorithm for finding a shattering set in a graph. Below we describe the algorithm and demonstrate its asymptotic optimality for random cubic graphs. Let $G\sim G_{n.3}$ be a random cubic graph. Consider the following algorithm:

{\em Algorithm Shatter, Phase I}
\begin{enumerate}
  \item Input: a graph $G$ and threshold $t$
  \item Find a Hamilton cycle $H$ in $G$
  \item Start from an arbitrary node $v_0$ and advance along $H$ creating a segment.
  \item When visiting a node $v$, it is incident with two edges on the cycle and a third edge $e$. If $e$ is the second edge in the segment going backward (in $H$), delete $v$.
\end{enumerate}

Each edge is seen once going forward and once going backward, and we delete nodes for half of the edges seen going backward, we remove exactly $1/4$ of the nodes. This is optimal as can be seen in Eq.\ \eqref{eq:expansion_bound}. A demonstration of Phase I is given in Figure \ref{fig:phase I}.

When Phase I is complete we are left with a tree $T$ of segments of $H$. These segments are with high probability of length $o(n)$ and are unicyclic. Shattering $T$ can be easily done by removing a center vertex of the tree, leaving at least two subtrees with at most half of the original number of segments in each. We continue in this manner until the graph is shattered, that is until every connected component is of size smaller than a predefined threshold. Summarizing we get:

{\em Algorithm Shatter, Phase II}
\begin{enumerate}
  \item Let $T$ be the tree of segments remaining after Phase I
  \item Until the maximal size of a connected component is smaller than $t$
\begin{enumerate}
  \item Find a center $v$ in the largest connected component
  \item Remove $v$
\end{enumerate}
\end{enumerate}
The vertices removed in Phase I and Phase II together form the shattering set.

The two phases are demonstrated in the following figures:
\begin{figure}
    \centering
    \begin{subfigure}[b]{0.45\textwidth}
        \includegraphics[width=\textwidth]{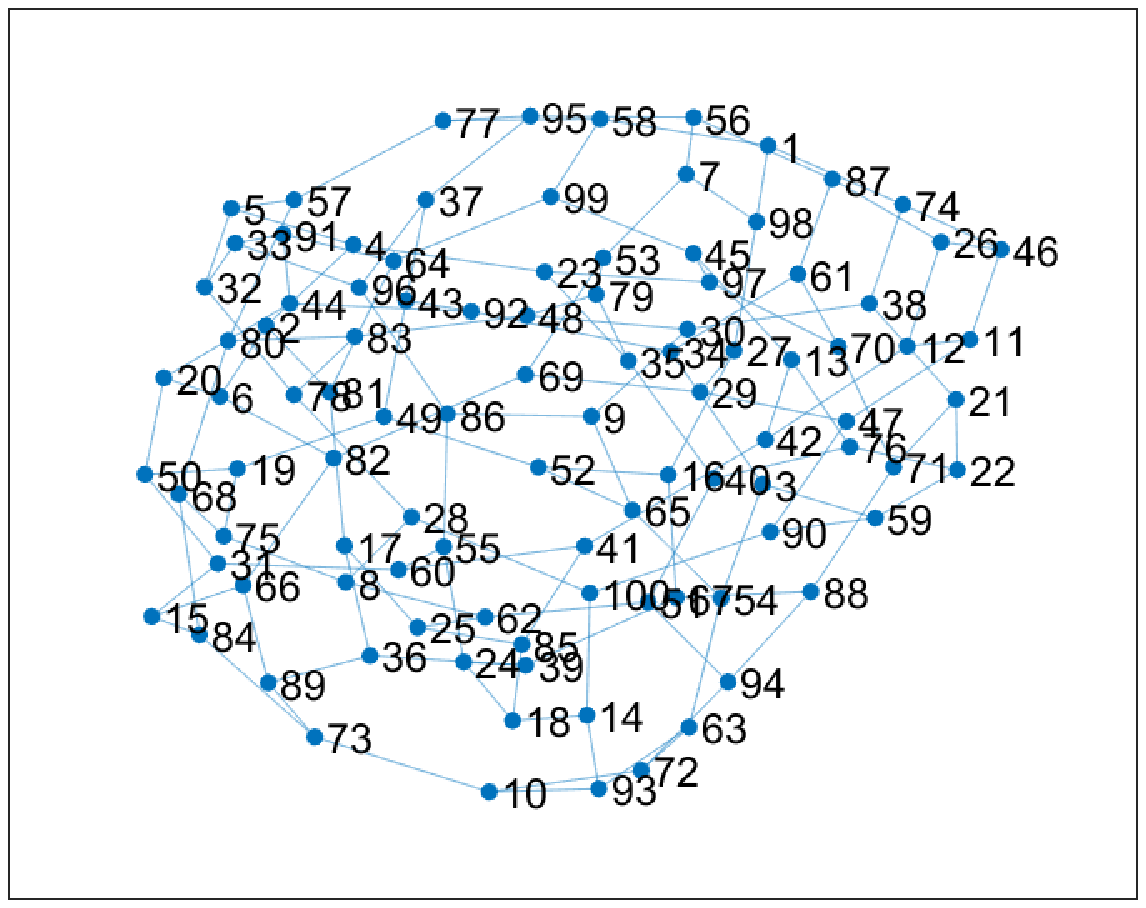}
        \caption{An instance of a random cubic graph with $n=100$ vertices}
        \label{sfig:random cubic}
    \end{subfigure}
    ~ %add desired spacing between images, e. g. ~, \quad, \qquad, \hfill etc.
    \begin{subfigure}[b]{0.45\textwidth}
        \includegraphics[width=\textwidth]{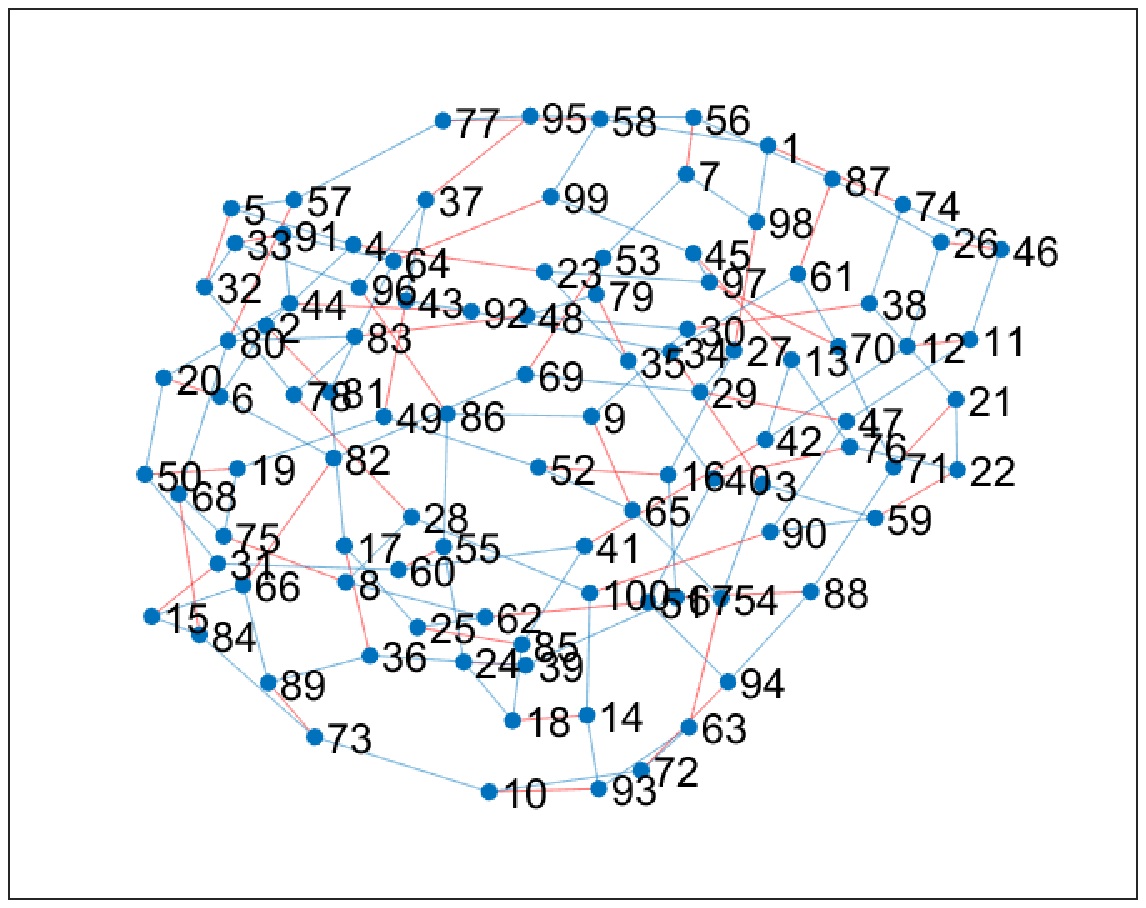}
        \caption{Same graph, edges out of a Hamilton cycle are colored red}
        \label{sfig:random cubic + ham}
    \end{subfigure}
    ~ %add desired spacing between images, e. g. ~, \quad, \qquad, \hfill etc.

    \begin{subfigure}[b]{0.45\textwidth}
        \includegraphics[width=\textwidth]{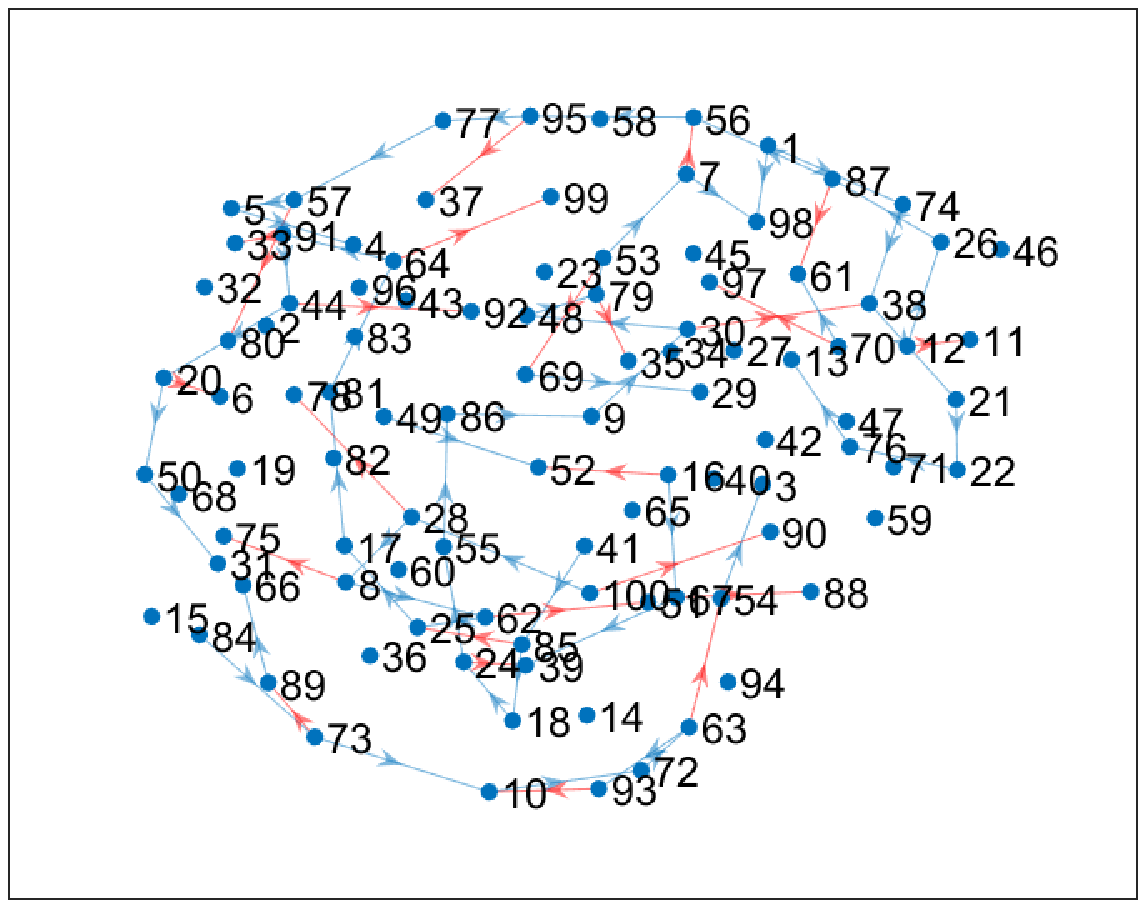}
        \caption{Same graph, after removal of every second forward going edge (end of phase I)}
        \label{sfig:random cubic after phase I}
    \end{subfigure}
    ~ %add desired spacing between images, e. g. ~, \quad, \qquad, \hfill etc.
    \begin{subfigure}[b]{0.45\textwidth}
        \includegraphics[width=\textwidth]{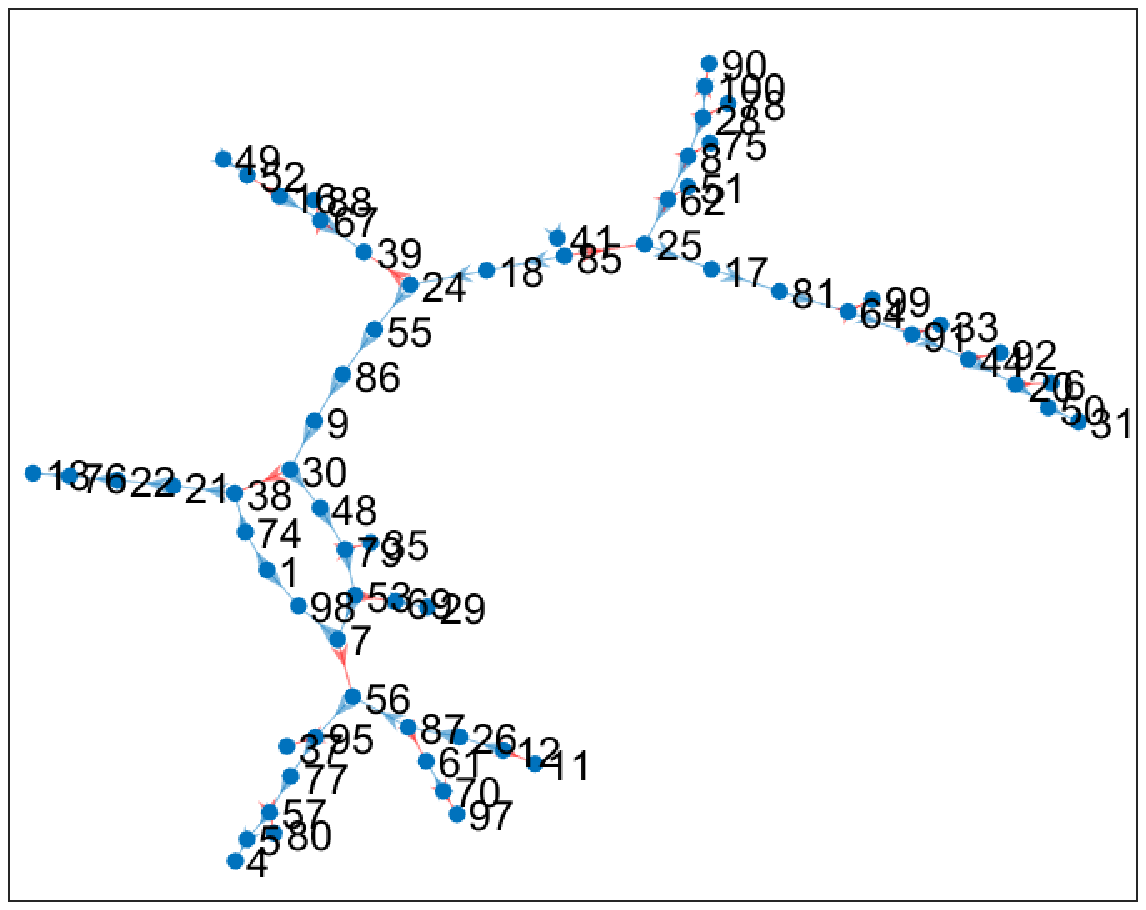}
        \caption{Giant component of the graph in Panel \ref{sfig:random cubic after phase I}, in tree layout.}
        \label{sfig:random cubic after phase I, tree layout}
    \end{subfigure}
    \caption{Phase I of Algorithm Shatter}\label{fig:phase I}
\end{figure}

Random cubic graphs are known to be Hamiltonian \cite{A:robinsom_&_wormald1992} and using a variant of ideas from the proofs from \cite{A:frieze_&_jerrum_&_molloy_&_robinson_&_wormald1996} we could get an algorithm finding such a cycle with high probability in time $O(n^{7/2})$. The rest of the algorithm (in both phases) runs in linear time. Preliminary results using a Branch and Bound approach suggest we may be able to reduce the running time of the Hamilton cycle finding module.

\begin{figure}
\includegraphics[width=0.95\textwidth]{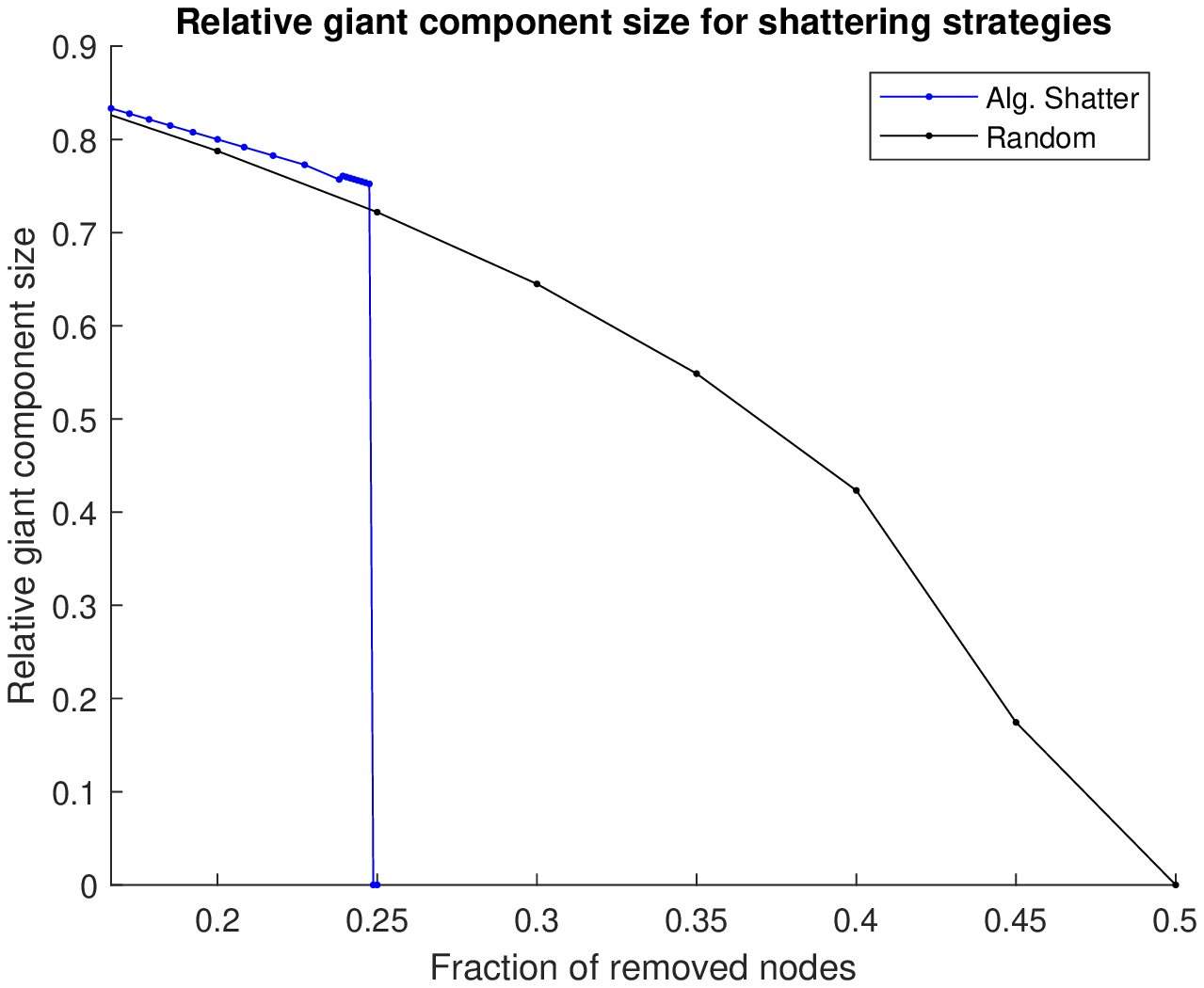}
\caption{Comparison of the performance of Algorithm Shatter vs Random Removal (which, in this case is also equivalent to Attack by Degree), for a random 3-regular graph with $10^5$ nodes.%
\label{fig:shatter vs deg for random cubic}}
\end{figure}

Notice that unlike removal by degree or by betweeness, which are local algorithms, the first phase of \emph{Shatter} finds a global structure in the network. This is our main insight --- we take advantage of the network randomness in order to find a global structure, then we use this structure to achieve high performance. In particular we get an asymptotically optimal constant for random cubic graphs (Fig.\ \ref{fig:shatter vs deg for random cubic}). While, as stated below, we can not guarantee optimality in all cases, we still believe this approach is favorable to local strategies, either passive or adaptive. This belief stems from the global nature of the problem, for which it is natural to suggest a global solution, and our attempt to optimize over a set rather than greedily point by point. Indeed, even when considering shattering to components of size one (i.e.\ finding an independent set), greedy algorithms are known to deliver poor results (in terms of approximation factor) when applied to random graphs \cite{A:grimmett_&_mcdiarmid1975}.

\captionsetup{width=0.9\textwidth}

For $d$-regular graphs with $d>3$ the same algorithm can be applied: tour along the Hamiltonian path, and remove the node at which a second edge going backward is observed. However, finding a Hamiltonian cycle in a $d$ regular graph with $d>3$ induces non-trivial correlations between the edges, and thus the performance of the algorithm is hard to evaluate. It is expected, however, that this algorithm will only retain an $O(1/k)$ fraction of the nodes, which is suboptimal.

For Erd\H{o}s-R\'enyi graphs $G(n,p)$ one may use a similar algorithm, with the following performance improvements:
\begin{enumerate}
  \item Every node of degree two can be replaced with an edge connecting its two neighbours. This is true, since long chains of length $O(n)$ occur with negligible probability in $G(n,p)$.
  \item Once the above step is performed, one can consider only the giant 3-core of the network, and perform {\em Algorithm Shatter} on it.
\end{enumerate}

%\begin{figure}
%\centering
%\includegraphics[]{}
%\caption{Random attack (which is also the degree based attack for a regular graph) vs the optimal
%attack for a random 3-regular graph with $n=3000$. The random threshold is $1/2$ whereas the optimal
%attack threshold is $1/4$.}
%\label{a107:fig:Ps}       % Give a unique label
%\vspace{-0.4cm}
%\end{figure}

\section{Summary}\label{Sec: Summary}
We provide bounds on the performance of optimal attack strategies on random networks, show that local strategies fail to achieve optimality even if used in adaptive manner and demonstrate an algorithm using global structure with optimal performance in certain situations. Our work draws the limits of feasibility for this well studied problem and shows that in some cases these limits are practically achievable. Besides their immediate value, our results may have broader implications. First, Algorithm Shatter is applicable whenever a long path or cycle may be found efficiently, e.g.\ when the network is based on a topological structure. Moreover, we see these results as an evidence for the ``global solution to global problem'' approach, and hope it will help in promoting this idea.

\begin{backmatter}

\section*{Abbreviations}
  Not applicable.

\section*{Availability of data and material}
  Not applicable.

\section*{Competing interests}
  The authors declare that they have no competing interests.

\section*{Funding}
  MK was partially supported by USA-Israel BSF grant 2014361, and by ISF grant 1261/17.
  This work was supported by the BIU Center for Research in Applied Cryptography and Cyber Security in conjunction with the Israel National Directorate in the Prime Minister’s office.

\section*{Author's contributions}
  All authors conducted the research. RC and SH wrote the manuscript. AH designed, coded and ran the simulations. All authors read and approved the final manuscript.

\section*{Acknowledgements}
  Not applicable.

\bibliographystyle{bmc-mathphys} % Style BST file (bmc-mathphys, vancouver, spbasic).
\bibliography{rgshtr}      % Bibliography file (usually '*.bib' )


\begin{thebibliography}{23}
\ifx \bisbn   \undefined \def \bisbn  #1{ISBN #1}\fi
\ifx \binits  \undefined \def \binits#1{#1}\fi
\ifx \bauthor  \undefined \def \bauthor#1{#1}\fi
\ifx \batitle  \undefined \def \batitle#1{#1}\fi
\ifx \bjtitle  \undefined \def \bjtitle#1{#1}\fi
\ifx \bvolume  \undefined \def \bvolume#1{\textbf{#1}}\fi
\ifx \byear  \undefined \def \byear#1{#1}\fi
\ifx \bissue  \undefined \def \bissue#1{#1}\fi
\ifx \bfpage  \undefined \def \bfpage#1{#1}\fi
\ifx \blpage  \undefined \def \blpage #1{#1}\fi
\ifx \burl  \undefined \def \burl#1{\textsf{#1}}\fi
\ifx \doiurl  \undefined \def \doiurl#1{\textsf{#1}}\fi
\ifx \betal  \undefined \def \betal{\textit{et al.}}\fi
\ifx \binstitute  \undefined \def \binstitute#1{#1}\fi
\ifx \binstitutionaled  \undefined \def \binstitutionaled#1{#1}\fi
\ifx \bctitle  \undefined \def \bctitle#1{#1}\fi
\ifx \beditor  \undefined \def \beditor#1{#1}\fi
\ifx \bpublisher  \undefined \def \bpublisher#1{#1}\fi
\ifx \bbtitle  \undefined \def \bbtitle#1{#1}\fi
\ifx \bedition  \undefined \def \bedition#1{#1}\fi
\ifx \bseriesno  \undefined \def \bseriesno#1{#1}\fi
\ifx \blocation  \undefined \def \blocation#1{#1}\fi
\ifx \bsertitle  \undefined \def \bsertitle#1{#1}\fi
\ifx \bsnm \undefined \def \bsnm#1{#1}\fi
\ifx \bsuffix \undefined \def \bsuffix#1{#1}\fi
\ifx \bparticle \undefined \def \bparticle#1{#1}\fi
\ifx \barticle \undefined \def \barticle#1{#1}\fi
\ifx \bconfdate \undefined \def \bconfdate #1{#1}\fi
\ifx \botherref \undefined \def \botherref #1{#1}\fi
\ifx \url \undefined \def \url#1{\textsf{#1}}\fi
\ifx \bchapter \undefined \def \bchapter#1{#1}\fi
\ifx \bbook \undefined \def \bbook#1{#1}\fi
\ifx \bcomment \undefined \def \bcomment#1{#1}\fi
\ifx \oauthor \undefined \def \oauthor#1{#1}\fi
\ifx \citeauthoryear \undefined \def \citeauthoryear#1{#1}\fi
\ifx \endbibitem  \undefined \def \endbibitem {}\fi
\ifx \bconflocation  \undefined \def \bconflocation#1{#1}\fi
\ifx \arxivurl  \undefined \def \arxivurl#1{\textsf{#1}}\fi
\csname PreBibitemsHook\endcsname

%%% 1
\bibitem{A:albert_&_jeong_&_barabasi2000}
\begin{barticle}
\bauthor{\bsnm{Albert}, \binits{R.}},
\bauthor{\bsnm{Jeong}, \binits{H.}},
\bauthor{\bsnm{Barab\'{a}si}, \binits{A.-L.}}:
\batitle{Error and attack tolerance of complex networks}.
\bjtitle{Nature}
\bvolume{406},
\bfpage{378}--\blpage{382}
(\byear{2000}).
doi:\doiurl{10.1016/0031-8914(72)90045-6}
\end{barticle}
\endbibitem

%%% 2
\bibitem{A:cohen_&_erez_&_ben-avraham_&_havlin2000}
\begin{barticle}
\bauthor{\bsnm{Cohen}, \binits{R.}},
\bauthor{\bsnm{Erez}, \binits{K.}},
\bauthor{\bsnm{ben-Avraham}, \binits{D.}},
\bauthor{\bsnm{Havlin}, \binits{S.}}:
\batitle{Resilience of the internet to random breakdown}.
\bjtitle{Phys. Rev. Lett.}
\bvolume{85}(\bissue{21}),
\bfpage{4626}--\blpage{4628}
(\byear{2000}).
doi:\doiurl{10.1103/PhysRevLett.85.4626}
\end{barticle}
\endbibitem

%%% 3
\bibitem{A:callaway_&_newman_&_strogatz_&_watts2000}
\begin{barticle}
\bauthor{\bsnm{Callaway}, \binits{D.S.}},
\bauthor{\bsnm{Newman}, \binits{M.E.J.}},
\bauthor{\bsnm{Strogatz}, \binits{S.H.}},
\bauthor{\bsnm{Watts}, \binits{D.J.}}:
\batitle{Network robustness and fragility: Percolation on random graphs}.
\bjtitle{Phys. Rev. Lett.}
\bvolume{85},
\bfpage{5468}--\blpage{5471}
(\byear{2000}).
doi:\doiurl{10.1103/PhysRevLett.85.5468}
\end{barticle}
\endbibitem

%%% 4
\bibitem{A:cohen_&_erez_&_ben-avraham_&_havlin2001}
\begin{barticle}
\bauthor{\bsnm{Cohen}, \binits{R.}},
\bauthor{\bsnm{Erez}, \binits{K.}},
\bauthor{\bsnm{ben-Avraham}, \binits{D.}},
\bauthor{\bsnm{Havlin}, \binits{S.}}:
\batitle{Breakdown of the internet under intentional attack}.
\bjtitle{Phys. Rev. Lett.}
\bvolume{86}(\bissue{16}),
\bfpage{3682}--\blpage{3685}
(\byear{2001}).
doi:\doiurl{10.1103/PhysRevLett.86.3682}
\end{barticle}
\endbibitem

%%% 5
\bibitem{A:magoni2003}
\begin{barticle}
\bauthor{\bsnm{Magoni}, \binits{D.}}:
\batitle{Tearing down the internets}.
\bjtitle{IEEE J.\ Sel.\ Areas Commun.}
\bvolume{21}(\bissue{6}),
\bfpage{949}--\blpage{960}
(\byear{2003}).
doi:\doiurl{10.1109/JSAC.2003.814364}
\end{barticle}
\endbibitem

%%% 6
\bibitem{A:paul_&_cohen_&_sreenivasan_&_havlin_&_stanley2007}
\begin{barticle}
\bauthor{\bsnm{Paul}, \binits{G.}},
\bauthor{\bsnm{Cohen}, \binits{R.}},
\bauthor{\bsnm{Sreenivasan}, \binits{S.}},
\bauthor{\bsnm{Havlin}, \binits{S.}},
\bauthor{\bsnm{Stanley}, \binits{H.E.}}:
\batitle{Graph partitioning induced phase transitions}.
\bjtitle{Phys. Rev. Lett.}
\bvolume{99}(\bissue{11}),
\bfpage{115701}
(\byear{2007}).
doi:\doiurl{10.1103/PhysRevLett.99.115701}
\end{barticle}
\endbibitem

%%% 7
\bibitem{Braunstein12368}
\begin{barticle}
\bauthor{\bsnm{Braunstein}, \binits{A.}},
\bauthor{\bsnm{Dall{\textquoteright}Asta}, \binits{L.}},
\bauthor{\bsnm{Semerjian}, \binits{G.}},
\bauthor{\bsnm{Zdeborov{\'a}}, \binits{L.}}:
\batitle{Network dismantling}.
\bjtitle{Proceedings of the National Academy of Sciences}
\bvolume{113}(\bissue{44}),
\bfpage{12368}--\blpage{12373}
(\byear{2016}).
doi:\doiurl{10.1073/pnas.1605083113}.
\arxivurl{https://www.pnas.org/content/113/44/12368.full.pdf}
\end{barticle}
\endbibitem

%%% 8
\bibitem{Ren6554}
\begin{barticle}
\bauthor{\bsnm{Ren}, \binits{X.-L.}},
\bauthor{\bsnm{Gleinig}, \binits{N.}},
\bauthor{\bsnm{Helbing}, \binits{D.}},
\bauthor{\bsnm{Antulov-Fantulin}, \binits{N.}}:
\batitle{Generalized network dismantling}.
\bjtitle{Proceedings of the National Academy of Sciences}
\bvolume{116}(\bissue{14}),
\bfpage{6554}--\blpage{6559}
(\byear{2019}).
doi:\doiurl{10.1073/pnas.1806108116}.
\arxivurl{https://www.pnas.org/content/116/14/6554.full.pdf}
\end{barticle}
\endbibitem

%%% 9
\bibitem{A:morone_&_makse2015}
\begin{barticle}
\bauthor{\bsnm{Morone}, \binits{F.}},
\bauthor{\bsnm{A.~Makse}, \binits{H.}}:
\batitle{Influence maximization in complex networks through optimal
  percolation}.
\bjtitle{Nature}
\bvolume{524},
\bfpage{65}--\blpage{68}
(\byear{2015}).
doi:\doiurl{10.1038/nature14604}
\end{barticle}
\endbibitem

%%% 10
\bibitem{PhysRevE.94.012305}
\begin{barticle}
\bauthor{\bsnm{Mugisha}, \binits{S.}},
\bauthor{\bsnm{Zhou}, \binits{H.-J.}}:
\batitle{Identifying optimal targets of network attack by belief propagation}.
\bjtitle{Phys. Rev. E}
\bvolume{94},
\bfpage{012305}
(\byear{2016}).
doi:\doiurl{10.1103/PhysRevE.94.012305}
\end{barticle}
\endbibitem

%%% 11
\bibitem{A:osat_&_faqeeh_&_radicchi2017}
\begin{barticle}
\bauthor{\bsnm{Osat}, \binits{S.}},
\bauthor{\bsnm{Faqeeh}, \binits{A.}},
\bauthor{\bsnm{Radicchi}, \binits{F.}}:
\batitle{Optimal percolation on multiplex networks}.
\bjtitle{Nature Communications}
\bvolume{8},
\bfpage{1540}
(\byear{2017}).
doi:\doiurl{10.1038/s41467-017-01442-2}
\end{barticle}
\endbibitem

%%% 12
\bibitem{CERQUETI2019320}
\begin{barticle}
\bauthor{\bsnm{Cerqueti}, \binits{R.}},
\bauthor{\bsnm{Ferraro}, \binits{G.}},
\bauthor{\bsnm{Iovanella}, \binits{A.}}:
\batitle{Measuring network resilience through connection patterns}.
\bjtitle{Reliability Engineering \& System Safety}
\bvolume{188},
\bfpage{320}--\blpage{329}
(\byear{2019}).
doi:\doiurl{10.1016/j.ress.2019.03.030}
\end{barticle}
\endbibitem

%%% 13
\bibitem{PhysRevE.91.022805}
\begin{barticle}
\bauthor{\bsnm{Chen}, \binits{P.-Y.}},
\bauthor{\bsnm{Cheng}, \binits{S.-M.}}:
\batitle{Sequential defense against random and intentional attacks in complex
  networks}.
\bjtitle{Phys. Rev. E}
\bvolume{91},
\bfpage{022805}
(\byear{2015}).
doi:\doiurl{10.1103/PhysRevE.91.022805}
\end{barticle}
\endbibitem

%%% 14
\bibitem{doi:10.1504/IJCEE.2018.088314}
\begin{barticle}
\bauthor{\bsnm{Ferraro}, \binits{G.}},
\bauthor{\bsnm{Iovanella}, \binits{A.}}:
\batitle{Clairvoyant targeted attack on complex networks}.
\bjtitle{International Journal of Computational Economics and Econometrics}
\bvolume{8}(\bissue{1}),
\bfpage{41}--\blpage{62}
(\byear{2018}).
doi:\doiurl{10.1504/IJCEE.2018.088314}.
\arxivurl{https://www.inderscienceonline.com/doi/pdf/10.1504/IJCEE.2018.088314}
\end{barticle}
\endbibitem

%%% 15
\bibitem{Gao2016UniversalRP}
\begin{barticle}
\bauthor{\bsnm{Gao}, \binits{J.}},
\bauthor{\bsnm{Barzel}, \binits{B.}},
\bauthor{\bsnm{Barab{\'a}si}, \binits{A.-L.}}:
\batitle{Universal resilience patterns in complex networks}.
\bjtitle{Nature}
\bvolume{530},
\bfpage{307}--\blpage{312}
(\byear{2016})
\end{barticle}
\endbibitem

%%% 16
\bibitem{RAMIREZMARQUEZ2018466}
\begin{barticle}
\bauthor{\bsnm{Ramirez-Marquez}, \binits{J.E.}},
\bauthor{\bsnm{Rocco}, \binits{C.M.}},
\bauthor{\bsnm{Barker}, \binits{K.}},
\bauthor{\bsnm{Moronta}, \binits{J.}}:
\batitle{Quantifying the resilience of community structures in networks}.
\bjtitle{Reliability Engineering \& System Safety}
\bvolume{169},
\bfpage{466}--\blpage{474}
(\byear{2018}).
doi:\doiurl{10.1016/j.ress.2017.09.019}
\end{barticle}
\endbibitem

%%% 17
\bibitem{HOSSEINI201647}
\begin{barticle}
\bauthor{\bsnm{Hosseini}, \binits{S.}},
\bauthor{\bsnm{Barker}, \binits{K.}},
\bauthor{\bsnm{Ramirez-Marquez}, \binits{J.E.}}:
\batitle{A review of definitions and measures of system resilience}.
\bjtitle{Reliability Engineering \& System Safety}
\bvolume{145},
\bfpage{47}--\blpage{61}
(\byear{2016}).
doi:\doiurl{10.1016/j.ress.2015.08.006}
\end{barticle}
\endbibitem

%%% 18
\bibitem{A:edwards_&_farr2001}
\begin{barticle}
\bauthor{\bsnm{Edwards}, \binits{K.}},
\bauthor{\bsnm{Farr}, \binits{G.}}:
\batitle{Fragmentability of graphs}.
\bjtitle{Journal of Combinatorial Theory, Series B}
\bvolume{82}(\bissue{1}),
\bfpage{30}--\blpage{37}
(\byear{2001}).
doi:\doiurl{10.1006/jctb.2000.2018}
\end{barticle}
\endbibitem

%%% 19
\bibitem{A:edwards_&_farr2008}
\begin{barticle}
\bauthor{\bsnm{Edwards}, \binits{K.}},
\bauthor{\bsnm{Farr}, \binits{G.}}:
\batitle{Planarization and fragmentability of some classes of graphs}.
\bjtitle{Discrete Mathematics}
\bvolume{308}(\bissue{12}),
\bfpage{2396}--\blpage{2406}
(\byear{2008}).
doi:\doiurl{10.1016/j.disc.2007.05.007}
\end{barticle}
\endbibitem

%%% 20
\bibitem{A:frieze_&_luczak1992}
\begin{barticle}
\bauthor{\bsnm{Frieze}, \binits{A.M.}},
\bauthor{\bsnm{{\L}uczak}, \binits{T.}}:
\batitle{On the independence and chromatic numbers of random regular graphs}.
\bjtitle{Journal of Combinatorial Theory, Series B}
\bvolume{54}(\bissue{1}),
\bfpage{123}--\blpage{132}
(\byear{1992}).
doi:\doiurl{10.1016/0095-8956(92)90070-E}
\end{barticle}
\endbibitem

%%% 21
\bibitem{A:robinsom_&_wormald1992}
\begin{barticle}
\bauthor{\bsnm{Robinson}, \binits{R.W.}},
\bauthor{\bsnm{Wormald}, \binits{N.C.}}:
\batitle{Almost all cubic graphs are {H}amiltoniann}.
\bjtitle{Random Structures \& Algorithms}
\bvolume{3}(\bissue{2}),
\bfpage{117}--\blpage{125}
(\byear{1992}).
doi:\doiurl{10.1002/rsa.3240030202}
\end{barticle}
\endbibitem

%%% 22
\bibitem{A:frieze_&_jerrum_&_molloy_&_robinson_&_wormald1996}
\begin{barticle}
\bauthor{\bsnm{Frieze}, \binits{A.M.}},
\bauthor{\bsnm{Jerrum}, \binits{M.}},
\bauthor{\bsnm{Molloy}, \binits{M.}},
\bauthor{\bsnm{Robinson}, \binits{R.W.}},
\bauthor{\bsnm{Wormald}, \binits{N.C.}}:
\batitle{Generating and counting hamilton cycles in random regular graphs}.
\bjtitle{Journal of Algorithms}
\bvolume{21}(\bissue{1}),
\bfpage{176}--\blpage{198}
(\byear{1996}).
doi:\doiurl{10.1006/jagm.1996.0042}
\end{barticle}
\endbibitem

%%% 23
\bibitem{A:grimmett_&_mcdiarmid1975}
\begin{barticle}
\bauthor{\bsnm{Grimmett}, \binits{G.R.}},
\bauthor{\bsnm{McDiarmid}, \binits{C.J.H.}}:
\batitle{On colouring random graphs}.
\bjtitle{Mathematical Proceedings of the Cambridge Philosophical Society}
\bvolume{77}(\bissue{2}),
\bfpage{313}--\blpage{324}
(\byear{1975}).
doi:\doiurl{10.1017/S0305004100051124}
\end{barticle}
\endbibitem

\end{thebibliography}

\newcommand{\BMCxmlcomment}[1]{}

\BMCxmlcomment{

<refgrp>

<bibl id="B1">
  <title><p>Error and attack tolerance of complex networks</p></title>
  <aug>
    <au><snm>Albert</snm><fnm>R</fnm></au>
    <au><snm>Jeong</snm><fnm>H</fnm></au>
    <au><snm>Barab\'{a}si</snm><fnm>AL</fnm></au>
  </aug>
  <source>Nature</source>
  <pubdate>2000</pubdate>
  <volume>406</volume>
  <fpage>378</fpage>
  <lpage>-382</lpage>
  <url>https://doi.org/10.1038/35019019</url>
</bibl>

<bibl id="B2">
  <title><p>Resilience of the Internet to Random Breakdown</p></title>
  <aug>
    <au><snm>Cohen</snm><fnm>R</fnm></au>
    <au><snm>Erez</snm><fnm>K</fnm></au>
    <au><snm>Avraham</snm><fnm>D</fnm></au>
    <au><snm>Havlin</snm><fnm>S</fnm></au>
  </aug>
  <source>Phys. Rev. Lett.</source>
  <publisher>American Physical Society</publisher>
  <pubdate>2000</pubdate>
  <volume>85</volume>
  <issue>21</issue>
  <fpage>4626</fpage>
  <lpage>-4628</lpage>
  <url>https://link.aps.org/doi/10.1103/PhysRevLett.85.4626</url>
</bibl>

<bibl id="B3">
  <title><p>Network Robustness and Fragility: Percolation on Random
  Graphs</p></title>
  <aug>
    <au><snm>Callaway</snm><fnm>DS</fnm></au>
    <au><snm>Newman</snm><fnm>MEJ</fnm></au>
    <au><snm>Strogatz</snm><fnm>SH</fnm></au>
    <au><snm>Watts</snm><fnm>DJ</fnm></au>
  </aug>
  <source>Phys. Rev. Lett.</source>
  <publisher>American Physical Society</publisher>
  <pubdate>2000</pubdate>
  <volume>85</volume>
  <fpage>5468</fpage>
  <lpage>-5471</lpage>
  <url>https://link.aps.org/doi/10.1103/PhysRevLett.85.5468</url>
</bibl>

<bibl id="B4">
  <title><p>Breakdown of the Internet under Intentional Attack</p></title>
  <aug>
    <au><snm>Cohen</snm><fnm>R</fnm></au>
    <au><snm>Erez</snm><fnm>K</fnm></au>
    <au><snm>Avraham</snm><fnm>D</fnm></au>
    <au><snm>Havlin</snm><fnm>S</fnm></au>
  </aug>
  <source>Phys. Rev. Lett.</source>
  <publisher>American Physical Society</publisher>
  <pubdate>2001</pubdate>
  <volume>86</volume>
  <issue>16</issue>
  <fpage>3682</fpage>
  <lpage>-3685</lpage>
  <url>https://link.aps.org/doi/10.1103/PhysRevLett.86.3682</url>
</bibl>

<bibl id="B5">
  <title><p>Tearing down the Internets</p></title>
  <aug>
    <au><snm>Magoni</snm><fnm>D.</fnm></au>
  </aug>
  <source>IEEE J.\ Sel.\ Areas Commun.</source>
  <publisher>IEEE</publisher>
  <pubdate>2003</pubdate>
  <volume>21</volume>
  <issue>6</issue>
  <fpage>949</fpage>
  <lpage>-960</lpage>
</bibl>

<bibl id="B6">
  <title><p>Graph Partitioning Induced Phase Transitions</p></title>
  <aug>
    <au><snm>Paul</snm><fnm>G</fnm></au>
    <au><snm>Cohen</snm><fnm>R</fnm></au>
    <au><snm>Sreenivasan</snm><fnm>S</fnm></au>
    <au><snm>Havlin</snm><fnm>S</fnm></au>
    <au><snm>Stanley</snm><fnm>HE</fnm></au>
  </aug>
  <source>Phys. Rev. Lett.</source>
  <publisher>American Physical Society</publisher>
  <pubdate>2007</pubdate>
  <volume>99</volume>
  <issue>11</issue>
  <fpage>115701</fpage>
  <url>https://link.aps.org/doi/10.1103/PhysRevLett.99.115701</url>
</bibl>

<bibl id="B7">
  <title><p>Network dismantling</p></title>
  <aug>
    <au><snm>Braunstein</snm><fnm>A</fnm></au>
    <au><snm>Dall{\textquoteright}Asta</snm><fnm>L</fnm></au>
    <au><snm>Semerjian</snm><fnm>G</fnm></au>
    <au><snm>Zdeborov{\'a}</snm><fnm>L</fnm></au>
  </aug>
  <source>Proceedings of the National Academy of Sciences</source>
  <publisher>National Academy of Sciences</publisher>
  <pubdate>2016</pubdate>
  <volume>113</volume>
  <issue>44</issue>
  <fpage>12368</fpage>
  <lpage>-12373</lpage>
  <url>https://www.pnas.org/content/113/44/12368</url>
</bibl>

<bibl id="B8">
  <title><p>Generalized network dismantling</p></title>
  <aug>
    <au><snm>Ren</snm><fnm>XL</fnm></au>
    <au><snm>Gleinig</snm><fnm>N</fnm></au>
    <au><snm>Helbing</snm><fnm>D</fnm></au>
    <au><snm>Antulov Fantulin</snm><fnm>N</fnm></au>
  </aug>
  <source>Proceedings of the National Academy of Sciences</source>
  <publisher>National Academy of Sciences</publisher>
  <pubdate>2019</pubdate>
  <volume>116</volume>
  <issue>14</issue>
  <fpage>6554</fpage>
  <lpage>-6559</lpage>
  <url>https://www.pnas.org/content/116/14/6554</url>
</bibl>

<bibl id="B9">
  <title><p>Influence maximization in complex networks through optimal
  percolation</p></title>
  <aug>
    <au><snm>Morone</snm><fnm>F</fnm></au>
    <au><snm>A. Makse</snm><fnm>H</fnm></au>
  </aug>
  <source>Nature</source>
  <pubdate>2015</pubdate>
  <volume>524</volume>
  <fpage>65</fpage>
  <lpage>-68</lpage>
  <url>https://doi.org/10.1038/nature14604</url>
</bibl>

<bibl id="B10">
  <title><p>Identifying optimal targets of network attack by belief
  propagation</p></title>
  <aug>
    <au><snm>Mugisha</snm><fnm>S</fnm></au>
    <au><snm>Zhou</snm><fnm>HJ</fnm></au>
  </aug>
  <source>Phys. Rev. E</source>
  <publisher>American Physical Society</publisher>
  <pubdate>2016</pubdate>
  <volume>94</volume>
  <fpage>012305</fpage>
  <url>https://link.aps.org/doi/10.1103/PhysRevE.94.012305</url>
</bibl>

<bibl id="B11">
  <title><p>Optimal percolation on multiplex networks</p></title>
  <aug>
    <au><snm>Osat</snm><fnm>S</fnm></au>
    <au><snm>Faqeeh</snm><fnm>A</fnm></au>
    <au><snm>Radicchi</snm><fnm>F</fnm></au>
  </aug>
  <source>Nature Communications</source>
  <pubdate>2017</pubdate>
  <volume>8</volume>
  <fpage>Articlenumber:1540</fpage>
  <url>https://www.nature.com/articles/s41467-017-01442-2</url>
</bibl>

<bibl id="B12">
  <title><p>Measuring network resilience through connection
  patterns</p></title>
  <aug>
    <au><snm>Cerqueti</snm><fnm>R</fnm></au>
    <au><snm>Ferraro</snm><fnm>G</fnm></au>
    <au><snm>Iovanella</snm><fnm>A</fnm></au>
  </aug>
  <source>Reliability Engineering \& System Safety</source>
  <pubdate>2019</pubdate>
  <volume>188</volume>
  <fpage>320</fpage>
  <lpage>329</lpage>
  <url>http://www.sciencedirect.com/science/article/pii/S0951832018310354</url>
</bibl>

<bibl id="B13">
  <title><p>Sequential defense against random and intentional attacks in
  complex networks</p></title>
  <aug>
    <au><snm>Chen</snm><fnm>PY</fnm></au>
    <au><snm>Cheng</snm><fnm>SM</fnm></au>
  </aug>
  <source>Phys. Rev. E</source>
  <publisher>American Physical Society</publisher>
  <pubdate>2015</pubdate>
  <volume>91</volume>
  <fpage>022805</fpage>
  <url>https://link.aps.org/doi/10.1103/PhysRevE.91.022805</url>
</bibl>

<bibl id="B14">
  <title><p>Clairvoyant targeted attack on complex networks</p></title>
  <aug>
    <au><snm>Ferraro</snm><fnm>G</fnm></au>
    <au><snm>Iovanella</snm><fnm>A</fnm></au>
  </aug>
  <source>International Journal of Computational Economics and
  Econometrics</source>
  <pubdate>2018</pubdate>
  <volume>8</volume>
  <issue>1</issue>
  <fpage>41</fpage>
  <lpage>62</lpage>
  <url>https://www.inderscienceonline.com/doi/abs/10.1504/IJCEE.2018.088314</url>
</bibl>

<bibl id="B15">
  <title><p>Universal resilience patterns in complex networks</p></title>
  <aug>
    <au><snm>Gao</snm><fnm>J</fnm></au>
    <au><snm>Barzel</snm><fnm>B</fnm></au>
    <au><snm>Barab{\'a}si</snm><fnm>AL</fnm></au>
  </aug>
  <source>Nature</source>
  <pubdate>2016</pubdate>
  <volume>530</volume>
  <fpage>307</fpage>
  <lpage>312</lpage>
</bibl>

<bibl id="B16">
  <title><p>Quantifying the resilience of community structures in
  networks</p></title>
  <aug>
    <au><snm>Ramirez Marquez</snm><fnm>JE</fnm></au>
    <au><snm>Rocco</snm><fnm>CM</fnm></au>
    <au><snm>Barker</snm><fnm>K</fnm></au>
    <au><snm>Moronta</snm><fnm>J</fnm></au>
  </aug>
  <source>Reliability Engineering \& System Safety</source>
  <pubdate>2018</pubdate>
  <volume>169</volume>
  <fpage>466</fpage>
  <lpage>474</lpage>
  <url>http://www.sciencedirect.com/science/article/pii/S0951832017303915</url>
</bibl>

<bibl id="B17">
  <title><p>A review of definitions and measures of system
  resilience</p></title>
  <aug>
    <au><snm>Hosseini</snm><fnm>S</fnm></au>
    <au><snm>Barker</snm><fnm>K</fnm></au>
    <au><snm>Ramirez Marquez</snm><fnm>JE</fnm></au>
  </aug>
  <source>Reliability Engineering \& System Safety</source>
  <pubdate>2016</pubdate>
  <volume>145</volume>
  <fpage>47</fpage>
  <lpage>61</lpage>
  <url>http://www.sciencedirect.com/science/article/pii/S0951832015002483</url>
</bibl>

<bibl id="B18">
  <title><p>Fragmentability of Graphs</p></title>
  <aug>
    <au><snm>Edwards</snm><fnm>K</fnm></au>
    <au><snm>Farr</snm><fnm>G</fnm></au>
  </aug>
  <source>Journal of Combinatorial Theory, Series B</source>
  <pubdate>2001</pubdate>
  <volume>82</volume>
  <issue>1</issue>
  <fpage>30</fpage>
  <lpage>-37</lpage>
  <url>http://www.sciencedirect.com/science/article/pii/S0095895600920185</url>
</bibl>

<bibl id="B19">
  <title><p>Planarization and fragmentability of some classes of
  graphs</p></title>
  <aug>
    <au><snm>Edwards</snm><fnm>K</fnm></au>
    <au><snm>Farr</snm><fnm>G</fnm></au>
  </aug>
  <source>Discrete Mathematics</source>
  <pubdate>2008</pubdate>
  <volume>308</volume>
  <issue>12</issue>
  <fpage>2396</fpage>
  <lpage>-2406</lpage>
  <url>http://www.sciencedirect.com/science/article/pii/S0012365X07003445</url>
</bibl>

<bibl id="B20">
  <title><p>On the Independence and Chromatic Numbers of Random Regular
  Graphs</p></title>
  <aug>
    <au><snm>Frieze</snm><fnm>AM</fnm></au>
    <au><snm>{\L}uczak</snm><fnm>T</fnm></au>
  </aug>
  <source>Journal of Combinatorial Theory, Series B</source>
  <pubdate>1992</pubdate>
  <volume>54</volume>
  <issue>1</issue>
  <fpage>123</fpage>
  <lpage>-132</lpage>
  <url>http://www.sciencedirect.com/science/article/pii/009589569290070E</url>
</bibl>

<bibl id="B21">
  <title><p>Almost all cubic graphs are {H}amiltoniann</p></title>
  <aug>
    <au><snm>Robinson</snm><fnm>RW</fnm></au>
    <au><snm>Wormald</snm><fnm>NC</fnm></au>
  </aug>
  <source>Random Structures \& Algorithms</source>
  <pubdate>1992</pubdate>
  <volume>3</volume>
  <issue>2</issue>
  <fpage>117</fpage>
  <lpage>-125</lpage>
  <url>http://onlinelibrary.wiley.com/wol1/doi/10.1002/rsa.3240030202/abstract</url>
</bibl>

<bibl id="B22">
  <title><p>Generating and Counting Hamilton Cycles in Random Regular
  Graphs</p></title>
  <aug>
    <au><snm>Frieze</snm><fnm>AM</fnm></au>
    <au><snm>Jerrum</snm><fnm>M</fnm></au>
    <au><snm>Molloy</snm><fnm>M</fnm></au>
    <au><snm>Robinson</snm><fnm>RW</fnm></au>
    <au><snm>Wormald</snm><fnm>NC</fnm></au>
  </aug>
  <source>Journal of Algorithms</source>
  <pubdate>1996</pubdate>
  <volume>21</volume>
  <issue>1</issue>
  <fpage>176</fpage>
  <lpage>-198</lpage>
  <url>http://dx.doi.org/10.1006/jagm.1996.0042</url>
</bibl>

<bibl id="B23">
  <title><p>On colouring random graphs</p></title>
  <aug>
    <au><snm>Grimmett</snm><fnm>GR</fnm></au>
    <au><snm>McDiarmid</snm><fnm>CJH</fnm></au>
  </aug>
  <source>Mathematical Proceedings of the Cambridge Philosophical
  Society</source>
  <pubdate>1975</pubdate>
  <volume>77</volume>
  <issue>2</issue>
  <fpage>313</fpage>
  <lpage>-324</lpage>
  <url>https://www.cambridge.org/core/journals/mathematical-proceedings-of-the-cambridge-philosophical-society/article/on-colouring-random-graphs/38719DFD65620CD94A379293A3E50B7F</url>
</bibl>

</refgrp>
} % end of \BMCxmlcomment

\end{backmatter}

\end{document}